\documentclass[a4paper,11pt]{article}
\usepackage{pos}

\usepackage{physics}
\usepackage{graphicx}
\usepackage{pgf}

\usepackage{varioref}
\usepackage[nameinlink]{cleveref}
\crefname{table}{table}{tables}
\Crefname{table}{Table}{Tables}
\crefname{figure}{figure}{figures}
\Crefname{figure}{Figure}{Figures}

\usepackage{acronym}
\newacro{EFT}[EFT]{effective field theory}
\acrodefplural{EFT}{effective field theories}
\newacro{CFT}[CFT]{conformal field theory}
\acrodefplural{CFT}{conformal field theories}
\newacro{UV}[UV]{ultraviolet}
\newacro{IR}[IR]{infrared}
\newacro{KK}[KK]{Kaluza-Klein}
\newacro{CKN}[CKN]{Cohen-Kaplan-Nelson}
\newacro{CEB}[CEB]{covariant entropy bound}

\newenvironment{eqaed}
    {\begin{equation}
    \begin{aligned}
    }
    { 
    \end{aligned}
    \end{equation}
	\ignorespacesafterend
    }

\title{String theory in the infrared}

\author*[a]{Ivano Basile}

\affiliation[a]{Max-Planck-Institut f\"{u}r Physik (Werner-Heisenberg-Institut),\\
Boltzmannstraße 8, 85748 Garching, Germany}

\emailAdd{ibasile@mpp.mpg.de}

\abstract{I briefly summarize a recent research program aiming to probe the landscape of low-energy phases of string theory from a global perspective. Borrowing conceptual lessons from the swampland program, I will discuss how the effective theories of gravity produced by low-energy string theory are far from generic; rather, their infrared data is connected by universal scaling relations which become non-trivial in species limits. In particular, a worldsheet analysis reveals that higher-derivative Wilson coefficients and the vacuum energy exhibit UV/IR relations which are invisible from the viewpoint of effective field theory, leading to parametric inequalities which take the form of holographic bounds. This lends a more solid theoretical support to swampland-motivated phenomenological scenarios, and to the broader hopes of extracting less direct, but empirically accessible, signatures of string theory from cosmological observations.}

\FullConference{%
  Corfu Summer Institute 2025 "School and Workshops on Elementary Particle Physics and Gravity"\\
  27 April - 18 September 2025\\
  Corfu, Greece
}

\begin{document}

\renewcommand{\hookAfterAbstract}{
    \par\bigskip\bigskip\bigskip
    Report number: MPP-2026-75
}

\maketitle

\section{Introduction}\label{sec:introduction}
    
\noindent While the high-energy properties of string theory make it a leading candidate for a quantum completion of gravity and other interactions, their direct signatures are outside experimental reach for the foreseeable future. Barring a gravitational cutoff scale near the threshold probed by accelerators, this is a generic issue of quantum gravity: the nature of the \ac{UV} completion of a given low-energy phase of gravity manifests itself via higher-derivative corrections to the (Wilsonian) effective action. The extreme \ac{IR} physics of our universe is well-described by Einstein-Maxwell theory\footnote{Assuming all neutrinos are massive. I also neglect the dark sector and the topological strong sector \cite{Bachmaier:2025hqg}, which (together with axions, if any) could play an important role in addressing some of the problems of the standard model.}
\begin{eqaed}\label{eq:wilsonian_action}
    S_\text{eff} = \int d^dx \, \sqrt{-g} \left(M_{{\text{Pl}}}^{d-2} \, R - \, F^2 + \mathcal{O}(R^2,F^4, \dots) + \dots \right)
\end{eqaed}
in $d=4$ spacetime dimensions, whose leading higher-derivative corrections can take several forms. In gravity, well-defined sharp observables are extracted from scattering amplitudes\footnote{Physically, this is due to the appearance of increasingly large black holes when probing spacetime at increasingly smaller sub-Planckian distances. Mathematically, there are no gauge-invariant local observables, and global (relational) observables become asymptotic observables once different spacetime topologies are included.}, and thus one can consider such corrections up to field redefinitions. In this case, the \ac{UV} contribution to pure graviton scattering is controlled by the cubic Goroff-Sagnotti contraction of three Riemann (or Weyl) tensors; for instance, its coefficient $a_{R^3}$ gives the leading tree-level contribution to the single-minus helicity amplitude \cite{Caron-Huot:2022ugt}
\begin{eqaed}\label{eq:single-minus_amplitude}
    \abs{\mathcal{A}_\text{grav}^{+++-}} = a_{R^3} \, \frac{stu}{2M_\text{Pl}^2} + \dots \, ,
\end{eqaed}
a particularly convenient process to study, since tree-level matter contributions do not appear due to angular-momentum selection rules \cite{Caron-Huot:2022ugt}. If the coefficient $a_{R^3}$ is large in Planck units, its contribution can overtake loop terms even at low energies, and produce potentially observable signatures as graviton non-Gaussianities in the inflationary spectrum \cite{Maldacena:2011nz} which can be reliable to all orders in the derivative expansion. As for lower-order corrections, at the four-derivative level there are electromagnetic-gravitational terms of the schematic form $RFF$ which contribute to graviton-photon scattering; once again, for some choice of helicities the leading two-derivative contribution vanishes.  Corrections of this type can also contribute to the UV-sensitivity of certain tidal perturbations of (near-)extremal black holes \cite{Horowitz:2023xyl, Horowitz:2024dch}. \newline

\noindent \paragraph{Field-theoretic vs quantum-gravitational effects.} In general, if $m \ll \Lambda_\text{QG}$, the origin of such Wilson coefficients is twofold \cite{Calderon-Infante:2025ldq}: they comprise a field-theoretic term, which arises by integrating out massive fields and scales with the appropriate power of the mass gap $m$ to new physics, and a quantum-gravitational term, which scales with the cutoff $\Lambda_\text{QG}$ above which no \ac{EFT} description is available\footnote{This quantum-gravitational cutoff is also known as the \emph{species scale} \cite{Dvali:2001gx, Veneziano:2001ah, Dvali:2007hz, Dvali:2007wp, Dvali:2009ks, Dvali:2010vm, Dvali:2012uq, Caron-Huot:2024lbf, ValeixoBento:2025iqu}, but not all its definitions are equivalent \cite{Blumenhagen:2023yws, Aoufia:2024awo, Basile:2024oms, Calderon-Infante:2025ldq}. Here I take the perspective of \cite{Caron-Huot:2024lbf, Bedroya:2024ubj} and define $\Lambda_\text{QG}$ as the minimal cutoff above which no \ac{EFT}, no matter in which spacetime dimension, can reliably describe the physics.}. In Planck units, the coefficient of a local operator of dimension $n$ in $d$ spacetime dimensions is thus expected to have the schematic form
\begin{eqaed}\label{eq:double_wilson}
    a_n = a_n^\text{EFT} \left(\frac{M_\text{Pl}}{m}\right)^{n-d} + a_n^\text{QG} \left(\frac{M_\text{Pl}}{\Lambda_\text{QG}}\right)^{n-2} .
\end{eqaed}
It is worth emphasizing that the absence of any other scalings is a \ac{UV}-sensitive matter, settling which requires going beyond \ac{EFT}. At any rate, up to unexpected cancellations, the above effects can hence only probe the desired scale $\Lambda_\text{QG}$ if all the dominant field-theoretic data is integrated in. In $d=4$ only terms of the form $\mathcal{O}(R^2)$ are an exception; in pure gravity none of them contributes to (perturbative) scattering amplitudes\footnote{The Gauss-Bonnet term, however, contributes to black-hole entropy even in $d=4$ \cite{Clunan:2004tb}, which should be visible non-perturbatively in the S-matrix due to black-hole production.}, but in Einstein-Maxwell theory they can contribute to mixed graviton-photon scattering \cite{Cano:2021tfs}. In this case, if the theory is invariant under electromagnetic duality, the analysis of \cite{Cano:2021tfs} shows that these terms (specifically, the squared Ricci tensor) are the only ones appearing up to field redefinitions. Another scenario in which the quantum-gravitational contribution to higher-derivative corrections dominant at low energies is if the $\mathcal{O}(R^2)$ Wilson coefficients depend on light scalar fields, perhaps participating in the dark sector \cite{Bedroya:2025fwh}. In this case, the Gauss-Bonnet term enters perturbative scattering amplitudes even in four dimensions. This scenario is realized in perturbative string theory, since the dilaton $\phi$ (whose asymptotic value is the logarithm $\log g_s$ of the string coupling constant) must be light in order to achieve parametrically (i.e. arbitrarily) weak couplings. \newline

\begin{figure}[!t]
    \centering  \includegraphics[width=0.35\linewidth]{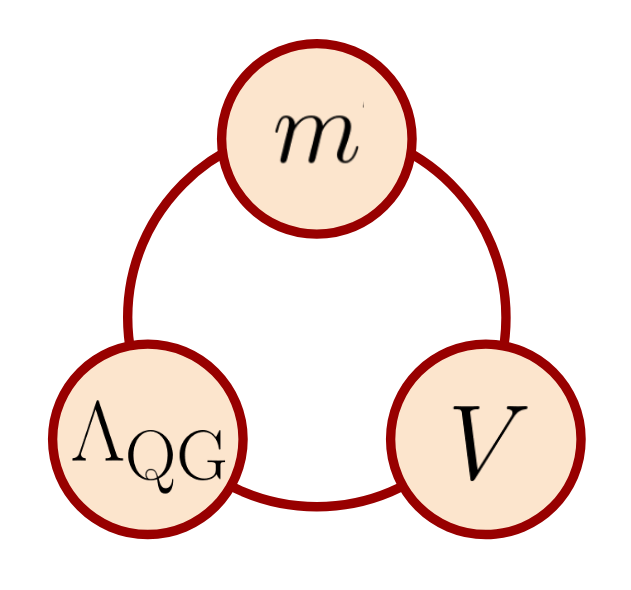}
    \caption{The three universal quantities in any gravitational \ac{EFT} are the mass gap $m$, the gravitational cutoff $\Lambda_\text{QG}$ and the vacuum energy $V$. I will review the parametric \ac{UV}/\ac{IR} relations connecting these quantities that emerge from string theory via the worldsheet approach.}
    \label{fig:triangle}
\end{figure}

\noindent \paragraph{\ac{UV}/\ac{IR} mixing.} Given the above considerations, it is clear that even in the best-case scenario the prospects of directly detecting effects driven by $\Lambda_\text{QG}$ are rather bleak, unless this scale is dramatically smaller than the Planck scale $M_\text{Pl}$\footnote{In this favorable scenario, or more precisely when there are no intermediate decompactification limits between the starting gravitational \ac{EFT} and its full \ac{UV}-completion, the latter is weakly coupled and is constrained to behave like string theory \cite{Camanho:2014apa, Bellazzini:2015cra, Caron-Huot:2016icg, Basile:2023blg, Bedroya:2024ubj, Herraez:2024kux}. For string scales accessible at particle accelerators, this leads to observable signatures such as those studied in \cite{Lust:2008qc}.}. A possible way out of this predicament is that the quantum-gravitational effects discussed above be unexpectedly related to low-energy data of the \ac{EFT}, such as the vacuum energy (or scalar potential) $V$ and gauge couplings $g$. This is a manifestation of \emph{\ac{UV}/\ac{IR} mixing}, an expected feature of gravity \cite{Cohen:1998zx, Bousso:1999xy} which abounds in string theory. These ideas underpin the program developed in \cite{Aoufia:2024awo, Basile:2024lcz, Aoufia:2026bau}, which I briefly review in the following. For the sake of brevity I will focus on the universal purely gravitational sector (and thus on closed strings), where the only low-energy data specifying the effective action is the vacuum energy density (in Planck units). The goal is then to connect higher-derivative Wilson coefficients to this quantity to uncover \ac{UV}/\ac{IR} relations. In particular, I will discuss relations connecting the mass gap $m$, the gravitational cutoff $\Lambda_\text{QG}$ and the vacuum energy (density) $V$, as depicted in \cref{fig:triangle}. By definition, to yield a gravitational \ac{EFT} with a parametric window of control, the ``Hubble'' parameter $\abs{V} \equiv M_\text{Pl}^{d-2} H^2$ must lie at the bottom of the hierarchy
\begin{eqaed}\label{eq:hierarchy}
    H \ll m \lesssim \Lambda_\text{QG} \lesssim M_\text{Pl} \, ,
\end{eqaed}
where $\lesssim$ is the logical negation of the asymptotic relation $\gg$\footnote{In this review it is always stipulated that parametric inequalities be defined within some limit in the space of parameters, moduli or theories, whichever the case may be.}. \newline

\begin{figure}[!t]
    \centering  \includegraphics[width=0.85\linewidth]{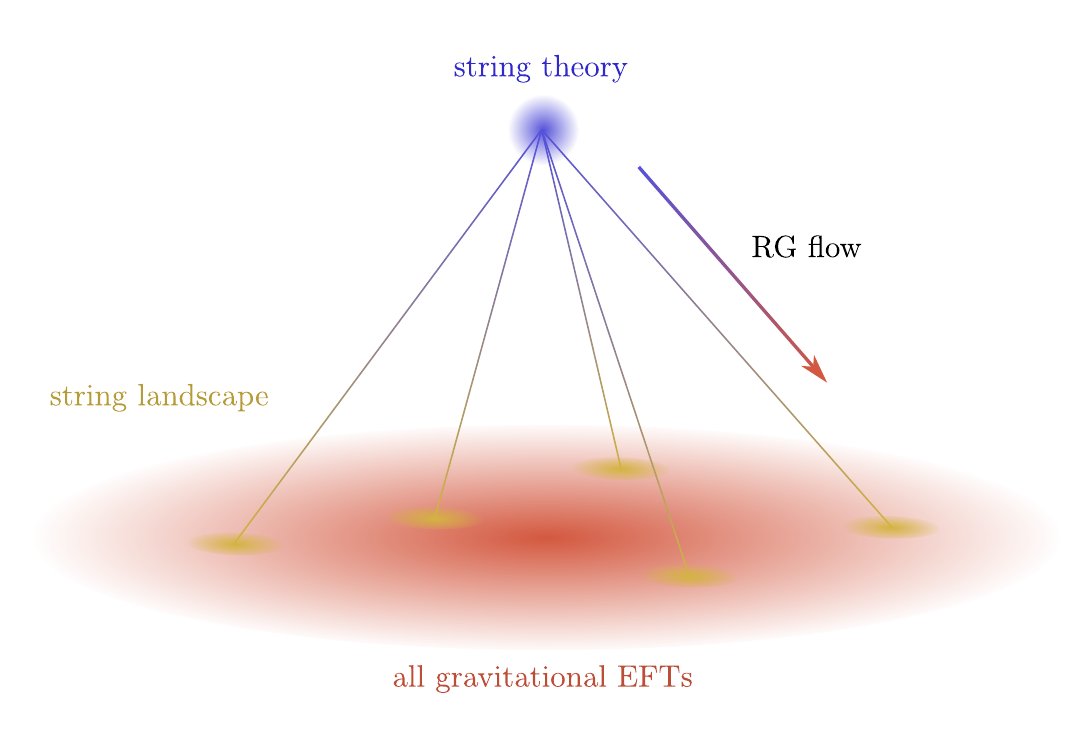}
    \caption{A depiction of the landscape of gravitational \acp{EFT} arising from string theory. The figure is meant to emphasize the sparse nature of the set of \ac{UV}-consistent \acp{EFT} relative to the set of all gravitational \acp{EFT}, which is the main driving force behind many aspects of the swampland program.}
    \label{fig:landscape}
\end{figure}

\noindent \paragraph{String theory in the infrared.} To pursue the search for \ac{UV}/\ac{IR} relations in string theory, the first task is to probe \emph{generic} features shared by its many low-energy phases, described by gravitational \acp{EFT} possibly connected by deformations or renormalization-group flows (see \cref{fig:landscape}). To this end, the worldsheet approach of \cite{Aoufia:2024awo, Basile:2024lcz, Aoufia:2026bau} encodes the data of a $d$-dimensional low-energy phase in a \ac{CFT} comprising the external spacetime sector and an internal (possibly non-geometric) sector. String perturbation theory then produces (the perturbative sector of) a gravitational \ac{EFT} (along with its high-energy completion) according to a map of the form
\begin{eqaed}\label{eq:string_worldsheet}
    \sigma(\mathcal{B}_d) \otimes \text{CFT}_\text{int}(t) \quad \longmapsto \quad \text{EFT}_d(\mathcal{B}_d, t) \, ,
\end{eqaed}
where $\text{CFT}_\text{int}(t)$ denotes the internal sector, namely a unitary (super)conformal theory with critical central charges with a unique vacuum and a discrete spectrum. For RNS-RNS constructions the critical central charges are $c = \overline{c} = 15 - 3d/2$, while in heterotic cases one is $26-d$. This \ac{CFT} can also have a conformal manifold, parametrized by local coordinates $t$\footnote{Strictly speaking, although I will keep this terminology throughout the review, the ensuing discussion only hinges on taking (species) limits in a family of \acp{CFT}.}. Such marginal deformations correspond to (pseudo-)moduli of the spacetime \ac{EFT}, namely (expectation values of) scalar fields whose tree-level potential vanishes. The symbol $\mathcal{B}_d$ denotes background data of the external macroscopic spacetime, i.e. profiles of spacetime fields (including the metric) that appear in the (superconformal) non-linear sigma model $\sigma$. When these profiles are uniformly weakly curved with respect to the string scale $\alpha' = M_s^{-2}$, low-energy physics is reliably captured by a gravitational \ac{EFT}. The upshot is that, in practice, one can restrict to the classical empty Minkowski background, since neighbouring backgrounds (including, hopefully, realistic cosmologies) fall within the same \ac{IR} phase. The strategy is then to compute Wilson coefficients from the string S-matrix and extract universal scaling relations between them. \newline

\noindent \paragraph{Summary of results.} The main findings of the one-loop analysis of \cite{Aoufia:2024awo, Basile:2024lcz}, subsequently extended in \cite{Aoufia:2026bau} including arguments for all-loop resummations, are the following:

\begin{itemize}
    \item \textbf{The quantum gravity cutoff.} The scale $\Lambda_\text{QG}$ is encoded in the internal worldsheet \ac{CFT} via the integral of its partition function $\mathcal{Z}_\text{int}$ over the moduli space of Riemann surfaces.
    
    \item \textbf{The emergent string dichotomy.} For generic (namely $O(1)$) values of the string coupling $g_s$ and worldsheet moduli $t$, all Wilson coefficients are $O(1)$ in Planck units. To achieve non-trivial scaling relations, including \ac{UV}/\ac{IR} mixing, one can either take $g_s \to 0^+$---resulting in asymptotically tensionless fundamental strings where $\Lambda_\text{QG}/M_\text{Pl} \ll 1$ is a positive power of $g_s$---or take a limit in the conformal manifold such that asymptotically $\Lambda_\text{QG} \ll M_\text{Pl}$ still. The latter \emph{results in a decompactification of (possibly emergent) extra dimensions}, supporting the emergent string conjecture \cite{Lee:2019wij} from the worldsheet perspective. Since this happens even for non-geometric internal sectors, in a sense this result can also be thought of as an extension of T-duality.
    
    \item \textbf{The vacuum energy and \ac{UV}/\ac{IR} relations.} In the above limits, the vacuum energy $V$ is dominated by the Casimir-like contribution $m^d$, where $m$ is the mass gap above the classical ground state (either the string scale $M_s$ or the \ac{KK} scale $m_\text{KK}$). In decompactification limits generically there is the additional contribution coming from dimensionally reducing the higher-dimensional vacuum energy, which is however too large to accommodate a reliable \ac{EFT}, unless the higher-dimensional string coupling is unacceptably small, and should thus vanish separately (barring unexpected cancellations, possibly involving higher-loop orders). This leads to parametric bounds reminiscent of holography, of the schematic form $\Lambda_\text{QG} \lesssim \abs{V}^{\alpha > 0}$ in Planck units.
\end{itemize}

\section{EFT from CFT: low-energy data from the worldsheet}\label{sec:low-energy_from_worldsheet}

\begin{figure}[!t]
    \centering  \includegraphics[width=0.65\linewidth]{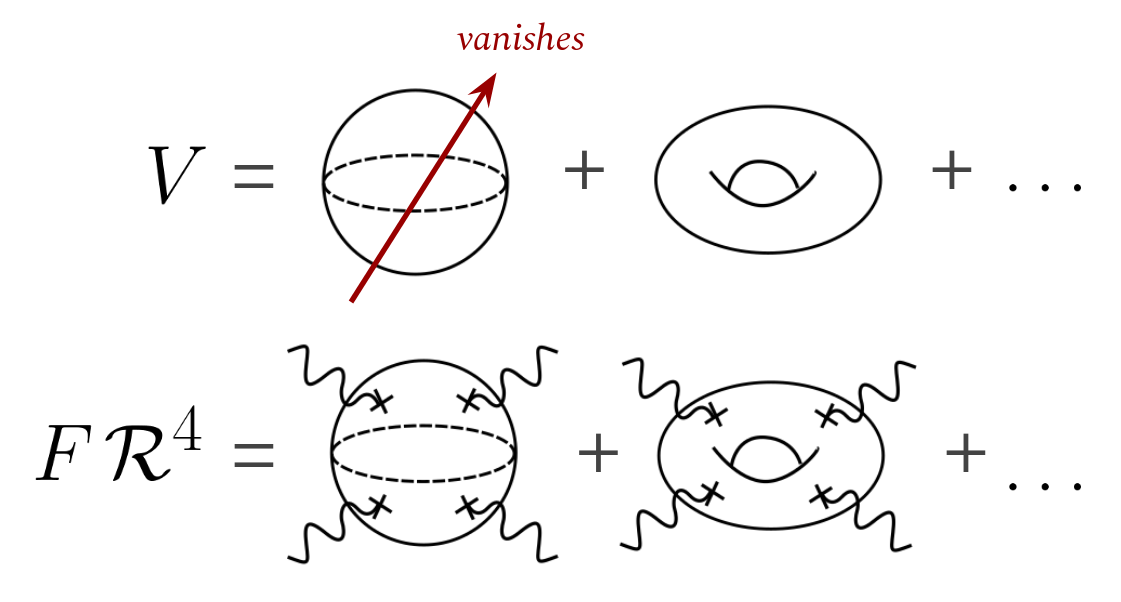}
    \caption{The (low-energy expansions of the) zero-point and four-point amplitudes, defined in \cref{eq:main_characters}, which extract the vacuum energy $V$ and the ``internal free energy'' $F$ (a probe of the gravitational cutoff $\Lambda_\text{QG}$ in closed-string perturbation theory. The tree-level (sphere) contribution to the vacuum amplitude vanishes, since it is related to the (on-shell) tree-level effective action on a flat background. This is an important further manifestation of how even a string-scale potential $V \sim M_s^d$ is driven by light species---in this case fundamental strings---rather than heavy non-perturbative objects (such as NS-branes).}
    \label{fig:amplitudes}
\end{figure}

\noindent As discussed above, the Wilson coefficients of interest are encoded in the low-energy expansion of string scattering amplitudes. This expansion involves several technical subtleties, including the need for \ac{IR} regularization in the presence of massless thresholds; the strategy of \cite{Aoufia:2024awo, Basile:2024lcz, Aoufia:2026bau} largely follows the blueprint laid out in \cite{Green:1999pv, Green:2008uj}. The main characters capturing the scaling properties of Wilson coefficients and vacuum energy are the (formally resummed) perturbative amplitudes
\begin{eqaed}\label{eq:main_characters}
    F(t,g_s) & = \sum_{h \geq 0} g_s^{2h-2} \int_{\mathcal{M}_h} d\mu_h(\tau) \, \mathcal{Z}^{(h)}_\text{int}(\tau;t) \, , \\
    V(t,g_s) & = \sum_{h \geq 0} g_s^{2h-2} \int_{\mathcal{M}_h} d\mu_h(\tau) \, Z^{(h)}_\text{ext}(\tau) \cdot \mathcal{Z}^{(h)}_\text{int}(\tau;t) \, ,
\end{eqaed}
where $\mathcal{Z}_\text{int}$ denotes the \emph{reduced} (or primary) partition function of the internal sector of the worldsheet \ac{CFT}, while $Z_\text{ext}$ denotes the remainder of the full partition function. In general, these objects are vector-valued in some space of sectors (such as spin structures or orbifold sectors), so that the dot product in \cref{eq:main_characters} is appropriately modular invariant\footnote{Similarly, in the expression for $F$ the functions $\mathcal{Z}_\text{int}$ ought to be contracted with a suitable constant vector, which in practice arises from the low-energy expansion of the external-spacetime contribution.}. Besides their formal nature as perturbative series, the expressions in \cref{eq:main_characters} hide a number of subtleties, discussed more extensively in \cite{Aoufia:2026bau}. Firstly, even in the absence of tachyons in the physical spectrum, non-supersymmetric vacua for which $V \neq 0$ induce tadpoles from some loop order in the expansion, which ought to be systematically subtracted along the lines of \cite{Fischler:1986ci, Kitazawa:2008hv} (see also \cite{Leone:2025mwo} for a recent review of worldsheet aspects of non-supersymmetric strings). Secondly, the integrals over the moduli spaces $\mathcal{M}_h$ of genus-$h$ Riemann surfaces against the Weil-Petersson measures $\mu_h$ ought to be replaced by their superconformal counterparts, which introduces significant mathematical complications. Finally, as anticipated, extracting local Wilson coefficients from loop-level amplitudes (which feature non-analytic terms) entails \ac{IR} regularization. However, none of these issues ends up being relevant for the scalings sought in \cite{Aoufia:2024awo, Basile:2024lcz}, as discussed in \cite{Aoufia:2026bau}, since they are unaffected by these modifications. For the same reason, the specific definition of $F$ in \cref{eq:main_characters} over the detailed structure of vertex-operator correlators pertaining to general scattering amplitudes does not change the overall scalings, which in the case of $F$ reflect those of a quartic curvature term of the form $\mathcal{O}(R^4)$. As such, as depicted in \cref{fig:amplitudes}, $V$ is computed by zero-point (i.e. vacuum) amplitudes, while $F$ is related to four-point amplitudes; it coincides with their low-energy behavior in some specific cases such as maximally supersymmetric type II vacua \cite{Green:1999pv, Green:2008uj}, but even more generally it correctly captures the scalings of the gravitational cutoff \cite{Aoufia:2024awo, Aoufia:2026bau}. \newline

\noindent In the following I will present the one-loop analysis of \cref{eq:main_characters} in species limits, starting from the punchline followed by the methodology. Then, I will discuss some arguments on higher-loop orders and the implications of the results for holography and phenomenology. \newline

\subsection{One-loop analysis: the gravitational cutoff and species limits} 

\begin{figure}[!t]
    \centering  \includegraphics[width=0.5\linewidth]{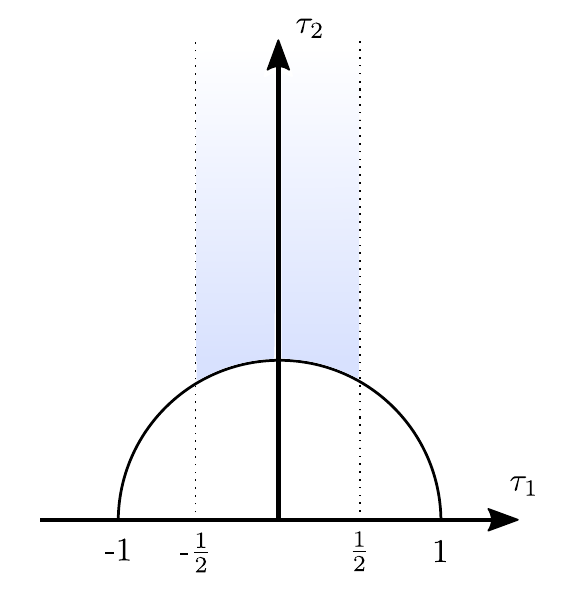}
    \caption{A depiction of the standard choice for the $\text{SL}(2,\mathbb{Z})$ fundamental domain $\mathcal{F}$, which parametrizes by a complex coordinate $\tau = \tau_1 + i \tau_2$ the moduli space $\mathcal{M}_1$ of conformal (equivalently complex) structures of a two-torus.}
    \label{fig:fundamental_domain}
\end{figure}

\noindent The tree-level contribution to $F$ and $V$ in \cref{eq:main_characters} is trivial (in particular, $V=0$ at tree level), reflecting the emergent-string limit driven by $g_s \to 0^+$. Thus, I will now focus on the one-loop contribution, where the moduli space $\mathcal{M}_1$ can be represented by the fundamental domain $\mathcal{F} = \mathbb{H}/\text{SL}(2,\mathbb{Z})$ parametrized by the complex coordinate $\tau = \tau_1 + i\tau_2$ (see \cref{fig:fundamental_domain}). As anticipated above, the interesting case for asymptotic scalings is a limit in the conformal manifold (parametrized by $t$) where the gravitational cutoff $\Lambda_\text{QG}$ vanishes in Planck units. In light of \cref{eq:double_wilson}, this means that $F$, and in particular its one-loop contribution $F_1$, should diverge. \newline

\noindent \paragraph{The existence of light species towers.} Due to modular invariance under $\text{SL}(2,\mathbb{Z})$, the general structure of the reduced partition function as a sum\footnote{The (sector-wise non-negative) degeneracies in \cref{eq:Z_int} are implicit in the sum. Alternatively, one can think of summing over individual states, including degenerate ones.} over spins $h - \overline{h} = j$ and conformal weights $h+\overline{h} = \Delta(t)$,
\begin{eqaed}\label{eq:Z_int}
    \mathcal{Z}_\text{int} = \tau_2^{\frac{c}{2}}\sum_{j,\Delta} e^{2\pi i j \tau_1} \, e^{-2\pi \Delta(t) \tau_2} \, ,
\end{eqaed}
implies that $F_1$ can only diverge if the spectral gap $\Delta_\text{gap}(t) \to 0^+$ in the limit, and is accompanied by an infinite tower of light states \cite{Aoufia:2024awo, Aoufia:2026bau}. To see this, consider splitting the \emph{canonical} partition function
\begin{eqaed}\label{eq:Z_split}
    Z_\text{can} \equiv \mathcal{Z}_\text{int}(\tau_1=0) \equiv Z_< + Z_>
\end{eqaed}
into a sum over conformal weights $\Delta$ that are above and below a non-zero threshold $\Delta_\text{th}$. In the following, for the sake of simplicity, I ignore the \ac{IR} regularization (which does not affect the proof) and the possibility of different sectors which mix under modular transformations, referring the reader to \cite{Aoufia:2026bau} for an explicit treatment of the general case. Letting $N$ be the number of states whose conformal weight is below $\Delta_\text{th}$, one can bound $Z_< \leq N \, \tau_2^{c/2}$. Moreover, for $\tau_2>1$,
\begin{eqaed}\label{eq:Z_bound}
    Z_> = \sum_{\Delta > \Delta_\text{th}} e^{-2\pi \Delta y + 2\pi \frac{\Delta}{y} - 2\pi \frac{\Delta}{y}} \leq e^{-2\pi \Delta_\text{th}\left( y - \frac{1}{y}\right)} \, Z'_> \, ,
\end{eqaed}
Where $Z'_>$ (resp. $Z'_<$) denotes the image of $Z_>$ (resp. $Z_<$) under the modular transformation $\tau_2 \mapsto 1/\tau_2$. Then, the modular invariance of \cref{eq:Z_split} entails that
\begin{eqaed}\label{eq:Z_modular_bound}
    Z_> \leq \frac{e^{-2\pi \Delta_\text{th}\left( \tau_2 - \frac{1}{\tau_2}\right)}}{1-e^{-2\pi \Delta_\text{th}\left( \tau_2 - \frac{1}{\tau_2}\right)}} \, N \, \tau_2^{\frac{c}{2}}
\end{eqaed}
for $\tau_2>1$, and an analogous bound for $\tau_2 < 1$. Since the integral over the fundamental domain $\mathcal{F}$ can be bounded by the integral of $Z_\text{can}$ on a strip where $\tau_2 \geq \sqrt{3}/2$, it follows that, taking a limit in the conformal manifold, $F_1$ cannot diverge unless $\Delta_\text{gap}(t) \to 0^+$ and $N \to +\infty$ for any $\Delta_\text{th} > \Delta_\text{gap}$ \cite{Aoufia:2024awo, Aoufia:2026bau}. \newline

\noindent \paragraph{Species limits in the conformal manifold are decompactifications.} In the language of the spacetime quantum-gravitational theory, the above is a \emph{species limit}, both definitionally (in the sense that the species scale vanishes in Planck units) and physically, since an infinite tower of light species becomes massless in Planck units. In turn, this implies that the limit lies at infinite distance in the conformal manifold \cite{Ooguri:2024ofs}, further supporting the link between species limits and infinite-distance limits outlined in \cite{Stout:2022phm}. Moreover, under conservative assumptions, the limiting \ac{CFT} is a sigma model on $\mathbb{R}^c$ \cite{Ooguri:2024ofs} accompanied by light \ac{KK} vertex operators. This already strongly suggests that any species limit in the worldsheet conformal manifold is a decompactification; the analysis of \cite{Aoufia:2024awo} further showed that
\begin{eqaed}\label{eq:volume}
    F_1 \overset{\Delta_\text{gap} \to 0^+}{\sim} \Delta_\text{gap}^{-\frac{c}{2}} = \left(\frac{M_\text{Pl}}{\Lambda_\text{QG}}\right)^{d-2} \equiv N_\text{species} \, ,
\end{eqaed}
which translates precisely to the scaling controlled by $\Lambda_\text{QG}$ for an $\mathcal{O}(R^4)$ term in a decompactification to ten dimensions, generalizing the results of \cite{Green:1999pv, Green:2008uj} to non-geometric settings. This strongly supports the conclusion that all such limits are decompactifications, and thus warrants the notation $\Delta_\text{gap}^{-\frac{c}{2}} \equiv \mathcal{V} = M_s^c\text{Vol}_c$ in terms of the (emergent) decompactifying volume in string units; this also follows by identifying the mass gap $m \equiv M_s \sqrt{\Delta_\text{gap}}$ with the \ac{KK} gap. For $c=2$, the appropriately \ac{IR}-regularized integral also produces the correct logarithmic term due to eight-dimensional massless thresholds. More complicated decompactification limits to other dimensions, analyzed in \cite{Aoufia:2024awo} and substantially generalized in \cite{Aoufia:2026bau}, also feature higher-dimensional logarithmic thresholds and the correct \ac{KK}-like terms of \cref{eq:double_wilson}. \newline

\subsection{One-loop analysis: the vacuum energy} 

\noindent The above discussion reviews how the gravitational cutoff $\Lambda_\text{QG}$, or species scale, weighing higher-derivative corrections controlled by the \ac{UV}-completion is encoded by the ``internal free energy'' $F$, as found in \cite{Aoufia:2024awo}. I now turn to the vacuum energy $V$, or scalar potential, which was studied with the same worldsheet approach in \cite{Basile:2024lcz}. Extending the methodology outlined below, which was introduced in \cite{Aoufia:2024awo} and subsequently refined in \cite{Aoufia:2026bau}, shows that in a species limit the one-loop contribution $V_1$ reads
\begin{eqaed}\label{eq:V_asymp}
    V_1 \overset{\Delta_\text{gap} \to 0^+}{\sim} a \, M_s^d \, \mathcal{V} + b \, m^d \, ,
\end{eqaed}
where the mass gap $m \equiv M_s \sqrt{\Delta_\text{gap}(t)} \lesssim M_s$ is bounded due to conformal symmetry \cite{Hellerman:2009bu, Hellerman:2010qd}. If $a = O(1)$ does not vanish, this closed-string contribution---which is simply the dimensional reduction of the higher-dimensional vacuum energy---yields a typical Hubble parameter $H \sim \Lambda_\text{QG}$, unless the (higher-dimensional) string coupling co-scales to zero rapidly enough; unacceptably so for a realistic model. Therefore, to obtain a realistic model one expects $a=0$ barring fine-tuning, and in fact for realistic values of all quantities the same should happen up to about 50 loops \cite{Basile:2024lcz}. This leaves us with the Casimir-like term in \cref{eq:V_asymp}; hence, any other contribution, including open-string sectors, should at most lead to the parametric inequality $V \gtrsim m^d$ \cite{Aoufia:2026bau}. Since generically $b=O(1)$, this parametric bound can only be violated for highly finely tuned numerical values of higher-loop contributions against the string coupling. \newline

\subsection{One-loop analysis: the methodology}

\noindent I will now sketch the arguments that lead to the above one-loop results. The starting point is the observation that reduced Narain partition functions $\mathcal{Z}_{c,c}$, which describe compactifications on a $c$-dimensional torus, satisfy the partial differential equation
\begin{eqaed}\label{eq:narain_asym_eq}
    \left(\Delta_{\text{Narain}} + w_c -\Delta_{\tau}\right) \mathcal{Z}_{c,c} = 0 \ , \quad w_c \equiv \frac{c}{2}\left(1-\frac{c}{2}\right) \, ,
\end{eqaed}
where $\Delta_{\text{Narain}}$ is the Laplacian on the Narain conformal manifold and $\Delta_\tau$ is the Laplacian on the fundamental domain of the worldsheet torus. Upon integrating over the latter (with a suitable modular-invariant regularization) against the modular-invariant measure $d\mu_1(\tau) = dxdy/y^2$, this leads to an eigenvalue Laplacian equation for $F_1$. The strategy followed in \cite{Aoufia:2024awo} is to replace the Narain \ac{CFT} with a general---possibly non-geometric---internal \ac{CFT} and seek a partial differential equation for $\mathcal{Z}_\text{int}$ along the lines of \cref{eq:narain_asym_eq}. While the existence of an exact such equation is hard to establish, it is possible to find a universal counterpart of \cref{eq:narain_asym_eq} which is valid \emph{asymptotically in species limits}. Specifically, parametrizing a curve in the conformal manifold by the coordinate $t \equiv \Delta_\text{gap}^{-1}$, there is a unique second-order differential operator $\mathcal{D}_c$ that replaces $\Delta_\text{Narain}$ in \cref{eq:narain_asym_eq}. In detail,
\begin{eqaed}\label{eq:asymptotic_diff_eq}
    \left( \mathcal{D}_c + w_c \right) \mathcal{Z}_{\text{int}} \overset{t \to +\infty}{\sim} \Delta_\tau \mathcal{Z}_{\text{int}} \ , \quad \mathcal{D}_{c} \equiv -t^2\partial_t^2 -(2-c)t \partial_t \, .
\end{eqaed}
A suitably \ac{IR}-regularized integral of both sides over the fundamental domain $\mathcal{F}$ can then be achieved by subtracting the real-analytic Eisenstein series $E_{c/2}$ for $c>2$, or its Kronecker limit $\widehat{E}_1$ for $c=1$ \cite{Green:2008uj, Benjamin:2021ygh, Aoufia:2024awo, Aoufia:2026bau}. The upshot is that the (regularized) integral satisfies
\begin{eqaed}\label{eq:integrated_ode}
    \mathcal{D}_c F_1 \overset{t \to +\infty}{\sim} - \, w_c F_1 \, ,
\end{eqaed}
whose dominant solution is indeed $F_1 \sim \mathcal{V}$. A more refined argument \cite{Aoufia:2024awo, Aoufia:2026bau} shows that the coefficient of this term is non-zero. In some (highly supersymmetric) settings, it is possible to show that the full Wilson coefficient, including non-perturbative contributions, satisfies an exact Laplacian equation analogous to \cref{eq:integrated_ode} \cite{Aoufia:2025ppe}. \newline

\noindent A similar argument for $V_1$ was developed in \cite{Basile:2024lcz}, then further generalized in \cite{Aoufia:2026bau}. One way to derive \cref{eq:V_asymp} is to use the harmonic decomposition of $\mathcal{Z}_\text{int}$ \cite{Benjamin:2021ygh}, whose zero-mode scales as $\mathcal{V}$, to obtain the first volume-like term in \cref{eq:V_asymp}. To obtain the Casimir-like term, one can write $\mathcal{Z}_\text{int} \equiv \mathcal{V} \left(A + \delta \mathcal{Z}(\tau;t) \right)$, where the constant $A$ yields the full zero-mode contribution and $\delta \mathcal{Z} \ll 1$ pointwise\footnote{This is due to the locality of the decompactification, which must restore a local higher-dimensional effective action. More general observables where the latter condition on $\delta \mathcal{Z}$ may not hold are discussed in \cite{Aoufia:2026bau}.}. Then, integrating against the external contribution $Z_\text{ext}$, the dominant contribution arises from its massless modes, encoded in the prefactor $y^{1-d/2-c/2}$, as discussed in more detail in \cite{Aoufia:2026bau}. Since this is an eigenfunction of $\Delta_\tau$, one can integrate by parts and obtain a modified differential equation for $V_1$ \cite{Basile:2024lcz}. Specifically, in the same notation as \cref{eq:asymptotic_diff_eq}, the extra term $\delta V_1$ satisfies
\begin{eqaed}\label{eq:dV_asymp_eq}
    \mathcal{D}_0 \, \frac{\delta V_1}{M_s^d \mathcal{V}} \overset{t \to +\infty}{\sim} w_{d+c} \, \frac{\delta V_1}{M_s^d \mathcal{V}} \, ,
\end{eqaed}
which leads to the Casimir-like scaling $\delta V_1 \sim m^d$.

\section{Higher-loop terms, holography and phenomenology}\label{sec:higher-loops}

\noindent The upshot of the above one-loop analysis is significant for theoretical purposes. However, even if a realistic model is well-approximated by a decompactification limit of perturbative string theory (possibly supplemented with non-perturbative ingredients), gauge couplings are expected to be set by the higher-dimension string coupling\footnote{An exception to this would be the ``little string'' scenario of \cite{Antoniadis:2001sw}, which however seems to require a large degree of fine tuning \cite{Basile:2024lcz}. See also \cite{Anchordoqui:2025nmb} for a related scenario in the context of the dark dimension \cite{Montero:2022prj}.}. This means that, at least parametrically speaking, one ought to check whether higher-loop contributions could spoil the above conclusions. At higher orders in the genus expansion, the structure of the moduli space of Riemann surfaces---especially its supersymmetric counterparts---increases in complexity rather quickly. \newline

\noindent Still, one can make progress by applying the above one-loop results to the effect that all species limits in the conformal manifold are decompactifications; hence, one expects that the dominant contribution to $F_h$ at each genus $h$ be given by the integral of the Narain partition function $\mathcal{Z}^{(h)}_{c,c}$. Much like its genus-one counterpart, this partition function satisfies a partial differential equation similar to \cref{eq:narain_asym_eq}, and affords an explicit expression as a lattice sum \cite{Maloney:2020nni}. For the purposes of this review, it is enough to observe that from either results one finds that the large-volume asymptotics of the genus-$h$ Narain lattice sum is $\mathcal{Z}^{(h)}_{c,c} \sim \mathcal{V}^h$. It follows that the genus expansion of $F$ in \cref{eq:main_characters} rearranges into the schematic form
\begin{eqaed}\label{eq:genus_expansion}
    F \overset{t \to +\infty}{\sim} \sum_{h \leq 0} g_s^{2h-2} \, c_h \, \mathcal{V}^h = \frac{1}{g_s^2}\sum_{h \leq 0} c_h \left(g_s^2\mathcal{V}\right)^h \, ,
\end{eqaed}
where the coefficients $c_h$ are expected to scale as $(2h)!$ up to power-like terms, as befits a resurgent expansion with D-brane instanton terms in its trans-perturbative sectors. This scaling in fact arises from the celebrated asymptotic results of Mirzakhani on moduli-space volumes \cite{Mirzakhani:2010pla}; as before, the supersymmetric version of the story is not expected to qualitatively affect \cref{eq:genus_expansion}. Similar conclusions apply to the volume-like terms in $V$, which generalize \cref{eq:V_asymp} to higher loops. \newline

\noindent \paragraph{The role of string dualities.} The main takeaway is that the perturbative expansion, at least for these terms, rearranges (and hopefully resums, upon including non-perturbative sectors) into a volume-like scaling times a function of the \emph{higher-dimensional string coupling} $g_s \sqrt{\mathcal{V}}$. As a result, insofar as string perturbation theory completes non-perturbatively, one can expect that volume-like scalings found in the one-loop analysis persist at least up to $O(1)$ values of the higher-dimensional string coupling. To probe strong couplings above this threshold, one must plausibly invoke dualities, possibly far from the (usually high-dimensional and highly supersymmetric) settings in which they are most established. One option is that the strongly coupled regime be (S-)dual to a weakly coupled sector. Another option is that the strong-coupling limit be a sector of M-theory in one dimension higher. Either way, the limit is driven by a decompactification or a unique, critical, asymptotically tensionless string, as befits the dichotomy of \cite{Lee:2019wij} now known as the emergent string conjecture. \newline

\begin{figure}[!t]
    \centering  \includegraphics[width=0.85\linewidth]{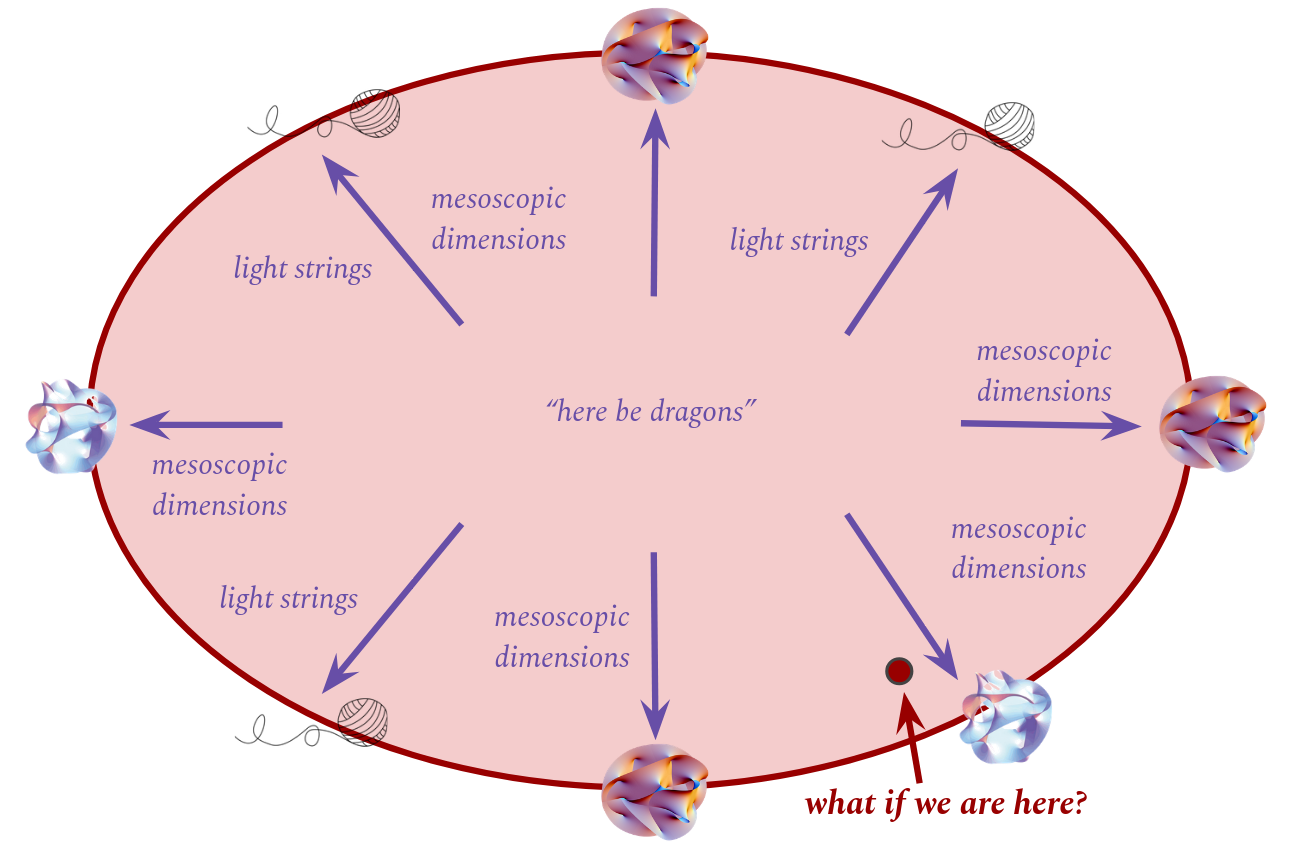}
    \caption{A cartoon of the possible species limits in string theory, in accord with the emergent string conjecture of \cite{Lee:2019wij}. The bounds in \cref{eq:holo_bound,eq:V_bound} suggest that our world might be located somewhat close to a decompactification limit. This was the conclusion of \cite{Montero:2022prj}, which led to the dark dimension proposal.}
    \label{fig:space}
\end{figure}

\noindent \paragraph{Implications for holography and applications to phenomenology.} The above considerations led us to conclude that the gravitational cutoff $\Lambda_\text{QG}$ is still related to the emergent decompactified volume according to \cref{eq:volume}, and that the vacuum energy $V$ satisfies the parametric bound
\begin{eqaed}\label{eq:V_bound}
    V \gtrsim m^d \, .
\end{eqaed}
Whether the mass gap $m$ arises from a decompactification or an emergent-string limit, in the \ac{EFT} settings at stake (where e.g. no ``sliding'' occurs \cite{Etheredge:2023odp, Raucci:2026fzp})
\begin{eqaed}\label{eq:species_scale}
    \frac{\Lambda_\text{QG}}{M_\text{Pl}} = \begin{cases}
        \ \left(\frac{m}{M_\text{Pl}}\right)^{\frac{c}{d+c-2}} \, , \quad \text{decompactification limit} \, , \\
        \ \ \ \frac{m}{M_\text{Pl}} \, , \quad \text{emergent-string limit} \, ,
    \end{cases}
\end{eqaed}
where the internal central charge $c$ of the decompactifying sector counts the number of (emergent) mesoscopic dimensions\footnote{More generally, for anisotropic decompactifications $c$ is replaced by an ``effective'' central charge $c_\text{eff}$ \cite{Castellano:2021mmx, Castellano:2022bvr}, as shown from the worldsheet approach in \cite{Aoufia:2026bau}.}. Hence, in any scenario $\Lambda_\text{QG} \lesssim M_\text{Pl} (m/M_\text{Pl})^{\frac{1}{d-1}}$, which together with \cref{eq:V_bound} leads to
\begin{eqaed}\label{eq:holo_bound}
    \Lambda_\text{QG} \lesssim M_\text{Pl} \left(\frac{\abs{V}}{M_\text{Pl}}\right)^{\frac{1}{d(d-1)}} \overset{\text{realistic model}}{\approx} 10^9 \text{ GeV} \, .
\end{eqaed}
From a theoretical perspective, these bounds may be interpreted as parametric avatars of holography. Indeed, on the one hand \cref{eq:V_bound} matches the \ac{CKN} bound \cite{Cohen:1998zx} applied to the \ac{EFT} with \ac{UV} cutoff $m$ and \ac{IR} scale the ``Hubble'' parameter $\abs{V} \equiv M_\text{Pl}^{d-2}H^2$. On the other hand, \cref{eq:holo_bound} matches the \ac{CEB} \cite{Bousso:1999xy} applied to $\Lambda_\text{QG}$ as \ac{UV} scale and $\abs{V}^\frac{1}{d}$ as \ac{IR} scale\footnote{These bounds have accrued recent interest in the context of cosmology \cite{Cribiori:2025oek} and in assessing the reliability of some string compactifications \cite{Cribiori:2026btb}.}. From a phenomenological perspective, these bounds suggest that our universe could be described by a (moderately weakly coupled) four-dimensional sector with some mesoscopic extra dimensions (see \cref{fig:space}). Observational data then constrains the standard model to be localized on a brane, and the (one \cite{Montero:2022prj}, or possibly two \cite{Anchordoqui:2025nmb} with some fine-tuning) extra dimension(s) to be of around micron size, leading to the dark dimension scenario proposed in \cite{Montero:2022prj} and recently developed in \cite{Bedroya:2025fwh} in light of recent cosmological data.

\section{Conclusions}\label{sec:conclusions}

\noindent In this brief review I attempted to summarize the salient lessons drawn from a global, swampland-inspired approach to the string landscape. The philosophy behind this approach is that the low-energy phases of string theory, while manifold, are very sparse relative to the set of all \acp{EFT}, as depicted in \cref{fig:landscape}. The requirement of a consistent \ac{UV}-completion poses stringent physical constraints on the landscape, which the community is gradually learning to wield in full force. \newline

\noindent Given the solid results on the stringy features of weakly coupled quantum gravity \cite{Camanho:2014apa, Bellazzini:2015cra, Caron-Huot:2016icg, Basile:2023blg, Bedroya:2024ubj, Herraez:2024kux}, the expectation of \ac{UV}/\ac{IR} mixing \cite{Cohen:1998zx, Bousso:1999xy} and the extreme smallness of the observed dark energy, it is natural to explore the pool of \ac{UV}-consistent gravitational \acp{EFT}---with an eye to its theoretical and phenomenological implications---from the vantage point of worldsheet string theory. This framework neatly encodes the set of low-energy phases in terms of the possible internal sectors of the worldsheet \ac{CFT}\footnote{Let me emphasize that, beyond the restriction to (trans-)perturbative physics, the RNS approach discussed in this review does not directly encompass non-trivial Ramond-Ramond backgrounds, orientifolds, or quantum vacua which do not exist at tree level. Nevertheless, in light of the considerations of \cref{sec:higher-loops}, I expect the overall conclusions to be robust enough not to be qualitatively affected by these more elaborate ingredients.}, allowing to formulate physical questions on (this corner of) the quantum-gravity landscape in much sharper mathematical terms. \newline

\noindent As a result, the \ac{IR} data of gravitational \acp{EFT} is manifestly expressed in terms of \ac{UV} data, and \ac{UV}/\ac{IR} mixing is captured by modular invariance and reflected in the universal scalings of \cref{eq:volume,eq:V_asymp} linking the mass gap $m$, the vacuum energy $V$ and the gravitational cutoff $\Lambda_\text{QG}$ (as well as gauge couplings and infinitely many other terms in the effective action \cite{Aoufia:2026bau}), as in \cref{fig:triangle}. Even including higher-order contributions, these findings resonate with holographic bounds and indicate tantalizing prospects for accessible observational evidence of indirect implications of quantum gravity for low-energy physics, indeed adhering to the spirit and goals of the swampland program. While the results obtained in \cite{Aoufia:2024awo, Basile:2024lcz, Aoufia:2026bau} are already encouraging in this respect, the wider program of understanding the universal low-energy implications of string theory, and their bearing---if any---on its uniqueness and necessity in quantum gravity, provides a rich arena for further research where exciting and deep questions can be formulated precisely and investigated via the powerful tools of \ac{CFT}. \newline 

\noindent Going beyond the worldsheet approach, the complete strong-coupling structure of \cref{eq:genus_expansion} hints at the key role played by string dualities, even beyond their most familiar incarnations in highly controlled settings, in reconstructing the full non-perturbative physics consistently with holography and related swampland ideas. Hence, following this reasoning to its extreme with a cavalier attitude, questions on the ultimate physical principles underpinning string theory could be deeply and surprisingly intertwined with questions on phenomenological mysteries such as the nature of dark energy, dark matter and the hierarchies found in our universe.

\section*{Acknowledgements}\label{sec:acknowledgements}

\noindent It is a pleasure to thank Christian Aoufia, Giorgio Leone, Matteo Lotito and Dieter L\"{u}st for related collaborations on which this review was based. I would also like to thank Giacomo Contri and Alessia Platania for useful comments on the manuscript and related discussions. This work was supported by the Origins Excellence Cluster and the German-Israel-Project (DIP) on Holography and the Swampland.

\end{document}